\documentclass[a4paper,11pt]{article}

\pdfoutput=1
\usepackage{jcappub}

\usepackage{graphicx}
\usepackage{longtable}
\usepackage{float}
\usepackage{dcolumn}
\usepackage{bm}
\usepackage{appendix}
\usepackage{multirow}
\usepackage{color}
\usepackage[utf8]{inputenc}
\usepackage{footmisc}
\usepackage{hyperref}

\usepackage{xcolor}
\usepackage{graphicx}
\usepackage{bm}
\usepackage{ulem}
\usepackage{multirow, array, float, tabularx}
\usepackage{amsmath}
\hypersetup{
	citecolor = red,
	linkcolor = blue,
	urlcolor = magenta
}

\newcommand{\be}{\begin{equation}}
	\newcommand{\ee}{\end{equation}}
\newcommand{\bea}{\begin{eqnarray}}
	\newcommand{\eea}{\end{eqnarray}}
\newcommand{\bes}{\begin{subequations}}
	\newcommand{\ees}{\end{subequations}}

\newcommand{\bc}{\begin{center}}
	\newcommand{\ec}{\end{center}}

\usepackage{wasysym}

\begin{document}

    \title{Stage IV CMB forecasts for warm inflation}

    \author[a]{F. B. M. dos Santos}\emailAdd{fbmsantos@on.br}
    \author[a]{G. Rodrigues}\emailAdd{gabrielrodrigues@on.br}    
    \author[a]{R. de Souza} \emailAdd{rayffsouza@on.br}
    \author[a]{J. S. Alcaniz}\emailAdd{alcaniz@on.br}

\affiliation[a]{Observatório Nacional, 20921-400, Rio de Janeiro - RJ, Brazil}

\abstract{We report forecast constraints on warm inflation in the light of future cosmic microwave background (CMB) surveys, with data expected to be available in the coming decade. These observations could finally give us the missing information necessary to unveil the production of gravitational waves during inflation, reflected by the detection of a non-zero tensor-to-scalar ratio crucial to the B-mode power spectrum of the CMB. We consider the impact of three future surveys, namely the CMB-S4, Simons Observatory, and the space-borne \textit{LiteBIRD}, in restricting the parameter space of four typical warm inflationary models in the context of a quartic potential, which is well motivated theoretically. We find that all three surveys significantly improve the models’ parameter space, compared to recent
results obtained with current \textit{Planck}+BICEP/Keck Array data. Moreover, the combination of ground-based and space-borne (SO+\textit{LiteBIRD} and CMB-S4+\textit{LiteBIRD}) surveys tightens the constraints so that we expect to distinguish even better warm inflation scenarios. This result becomes
clear when we compare the models’ predictions with a $\Lambda$CDM+$r$ forecast, compatible with $r=0$, in which one of them already becomes excluded by data.}

	\maketitle

\section{Introduction}\label{sec1}

Measurements of the cosmic microwave background (CMB) opened a new window of observation to the early universe \cite{WMAP:2012nax,Planck:2018nkj,Planck:2018vyg,BICEP:2021xfz}, not only providing tight constraints on cosmological parameters, but also allowing us to directly probe the hypothesis of inflation at very early times \cite{Guth:1980zm,Starobinsky:1980te,Linde:1981mu}, which solves a series of problems related to the standard cosmological model. In addition, inflation provides a mechanism to generate the perturbations that dictate the matter distribution at present times, as well as a connection between a new scalar field, the \textit{inflaton}, and particle production through the reheating of the universe \cite{Albrecht:1982mp,Abbott:1982hn,Kofman:1994rk,Kofman:1997yn}.

Although the current CMB data provide substantial hints at the occurrence of the so-called slow-roll inflationary period, an important piece is still missing: tensor perturbations generated during inflation can produce a gravitational wave background of primordial origin, which is reflected in the CMB as B-mode polarization. Additional sources of B modes are also provided by weak gravitational lensing, which changes polarization from E modes to B ones. Lensing constitutes the dominant part in the B-mode power spectrum, while fluctuations of primordial origin are only relevant at large scales. Lensing B modes were recently detected by the BICEP/Keck Array observatories \cite{BICEP2:2015xme,BICEP2:2018kqh,BICEP:2021xfz}, such that the tensor-to-scalar ratio $r$ (the ratio between amplitudes of tensor and scalar perturbations) is now constrained as $r<0.036$ (see also \cite{Tristram:2020wbi,Tristram:2021tvh}), enough to exclude many classes of well-motivated inflationary models.

As the value of $r$ gets lower, it is important to determine whether we have a realization of the universe consistent with $r=0$, or if there really is an inflationary  stochastic gravitational wave background that may be detectable. That is one of the goals of future CMB experiments, such as \textit{LiteBIRD}  \cite{Matsumura:2013aja,Hazumi:2019lys,LiteBIRD:2022cnt}, CMB-S4 \cite{CMB-S4:2016ple,CMB-S4:2020lpa}, and the Simons Observatory (SO) \cite{SimonsObservatory:2018koc}. When properly combined, they are projected to estimate $r$ with high precision, thus being expected to finally give us one of the more conclusive evidence for or against inflation.

Models in which $r$ can be small are well discussed in the literature. In particular, the idea of warm inflation (WI) \cite{Berera:1995ie,Maia:1999yt,Hall:2003zp,Berera:2023liv,Kamali:2023lzq}, where the inflaton interacts with radiation fields in a thermal bath, becomes an interesting possibility. Additionally, models which are currently excluded by observations e.g. the one given by a quartic potential, are now viable within the formalism. This comes at the cost of considering a suitable form for the dissipation coefficient $\Upsilon$ that contains the details of the interactions and usually is dependent on the temperature of the thermal bath. Over the past years, many realizations of the scenario, both at the fundamental and phenomenological level have been studied \cite{Berera:2008ar,Bastero-Gil:2010dgy,Benetti:2016jhf,Bastero-Gil:2017wwl,Arya:2017zlb,Motaharfar:2018zyb,Arya:2018sgw,Bastero-Gil:2019gao,Das:2019acf,Kamali:2019xnt,Benetti:2019kgw,Berera:2019zdd,Das:2020xmh,Santos:2022exm,Arya:2022xzc,Arya:2023pod,Ballesteros:2023dno,Kumar:2024hju}, in which it is possible to establish reasonable constraints with current CMB data. In a recent work, some of us have obtained updated constraints on the quartic model for four different dissipation coefficients \cite{Santos:2024pix}, all of which exhibited a preference for lower values of $r$, which is expected, but may vary according to the model considered. 

A useful manner to assess the capability of future experiments to constrain cosmological models is the forecasting procedure, which allows us, given information about the instrument and its limitations, to estimate the constraining power of many different scenarios. In this work, we investigate the constraints given only by future CMB data in a way that provides a more direct comparison with the current constraints obtained by the \textit{Planck}+BICEP/Keck Array. In our approach, warm inflationary models can depend on only one additional parameter, the ratio between the dissipation coefficient $\Upsilon$ and the rate of
cosmic expansion $H$, defined as  $Q\equiv\Upsilon/3H$, at the chosen pivot scale $k_\star$. Constraints on this parameter directly reflect restrictions on $r$, a derived parameter. Thus, we consider the individual impact of CMB-S4 and SO while performing a combined analysis with \textit{LiteBIRD}.

The work is organized as follows. In section \ref{sec2} we review recent results on the phenomenology of warm inflation models when confronted with the current CMB data. Such results will be the basis for our forecasts. In section \ref{sec3}, we describe the forecasting procedure along with the mock likelihoods to be used in our analysis. Section \ref{sec4} presents and discusses our main results. Finally, we present our conclusions in section \ref{sec5}.

\section{Current CMB constraints on the quartic warm inflation model}\label{sec2}

In this section, we review the formalism of warm inflation, as well as the results obtained in \cite{Santos:2024pix}.

\subsection{The warm inflation formalism}

The warm inflation idea proposed in \cite{Berera:1995ie} establishes that the slow-roll inflationary period was sustained by a thermal bath caused by the dissipation of the inflaton field into radiation. In a general manner, this is modeled by the presence of a dissipation coefficient $\Upsilon$ that carries the microphysics of the model in the scalar field equation of motion:
\begin{gather}
\ddot\phi + 3H\dot\phi + V_{,\phi} = -\Upsilon\dot\phi
\end{gather}
where $H=\frac{\dot a}{a}$ is the Hubble parameter, with a dot denoting a derivative with respect to cosmic time. Given the presence of the term on the right-hand side, the friction to the inflaton's motion is then characterized by a combination of the Hubble expansion and the dissipation rate. It is then convenient to define a parameter $Q$ as 
\begin{equation}
Q\equiv\frac{\Upsilon}{3H},
\end{equation}
that helps us to identify the dissipative regime in which inflation takes place. Specifically, $Q\ll 1$ characterizes the \textit{weak} dissipative regime, while the \textit{strong} regime happens for $Q\gg 1$.

The ratio $Q$ has a great impact on the determination of the slow-roll regime, given that the usual slow-roll parameters $\epsilon_V$ and $\eta_V$ are redefined as
\begin{equation}
\epsilon_w = \frac{\epsilon_V}{1+Q}, \quad \eta_w = \frac{\eta_V}{1+Q},
\end{equation}
meaning that even for $\epsilon_V,|\eta_V|>1$, inflation in the strong regime ($Q>1$) can lead to successful slow-roll inflation, as $\epsilon_w,|\eta_w|$ can still be lower than one.

Assuming rapid thermalization, warm inflation is achieved whenever $T/H>1$, where $T$ is the temperature of the bath. By the end of inflation, the radiation energy density can be comparable to that of the inflaton, thereby allowing the universe to smoothly transit into a radiation-dominated era. Due to the dissipative effects, the physics at the perturbative level are no longer the same. The primordial curvature power spectrum is modified as \cite{Graham:2009bf,Bastero-Gil:2009sdq,Bastero-Gil:2011rva}

\begin{equation} \label{9}
\Delta^2_{\mathcal{R}}(k_\star)  = \mathcal{P}_{\mathcal{R},c}\left(1 + 2n_{BE,\star} + \frac{2\sqrt{3}\pi Q_\star}{\sqrt{3+4\pi Q_\star}}\frac{T_\star}{H_\star}\right)G(Q_\star),
\end{equation}
while the tensor spectrum retains its usual form, $\Delta^2_T(k_\star) = \frac{2H_\star^2}{\pi^2 M_p^2}$. Here, $\mathcal{P}_{\mathcal{R},c} \equiv \left(\frac{H_\star^2}{2\pi\dot\phi_\star}\right)^2$ and $n_{BE,\star}=\frac{1}{e^{H_\star/T_\star}-1}$, where the subscript `$\star$' corresponds to quantities measured at the horizon crossing of the CMB pivot scale. $G(Q_\star)$ is a function determined numerically that accounts for interactions between inflaton and radiation perturbations, being usually more relevant in the strong dissipation regime. 

The spectral index $n_s$ and tensor-to-scalar ratio $r$ are still determined in the usual manner:
\begin{equation}
n_s-1 = \frac{d\ln\Delta^2_{\mathcal{R}}}{d\ln k}, \quad r=\frac{\Delta^2_T}{\Delta^2_\mathcal{R}}.
\end{equation}
However, the differences shown in \ref{9} reflect drastic changes in how the predictions for $n_s$ and $r$ will be when compared to CMB constraints. For instance, while the spectral index can be either well inside the current \textit{Planck} constraints or become blue/red-tilted, usually, the tensor-to-scalar ratio can be very low, down to $r\sim 10^{-5}$, or, in more extreme cases, when the very strong regime takes place at $Q_\star\sim 800$, $r$ can be as low as $r\sim 10^{-30}$ \cite{Das:2020xmh}.

\subsection{The quartic model}

In recent work \cite{Santos:2024pix}, some of us have performed an updated analysis on the quartic inflation model, given by the potential
\begin{equation}
V(\phi)=\frac{\lambda}{4}\phi^4,
\label{eq:2.6}
\end{equation}
which is currently excluded by CMB data in the standard cold inflation scenario. On the other hand, in the warm inflationary picture, concordance with CMB constraints is possible, allowing a direct comparison with the data in order to perform a parameter infererence. This was recently done in \cite{Arya:2017zlb,Arya:2018sgw,Bastero-Gil:2017wwl}; however, in our work, we have collected some of the most studied forms for $\Upsilon$ and have performed a Bayesian statistical comparison when B-mode polarization CMB data from the BICEP/Keck Array (BK18) telescopes is considered \cite{BICEP:2021xfz}, for models displayed in Table \ref{tab:1}. These are based on an earlier study done in \cite{Motaharfar:2018zyb}, in which the possibilities for the predictions on the $n_s-r$ plane and the swampland conjectures were discussed. In general, they can arise from a parametrization of $\Upsilon$, expressed as:
\begin{equation}
    \Upsilon=C_\Upsilon T^p \phi^c M^{1-p-c},
\end{equation}
where $C_\Upsilon$ and $M$ are constants.

\begin{figure*}[t]
\centering
\includegraphics[width=0.7\columnwidth]{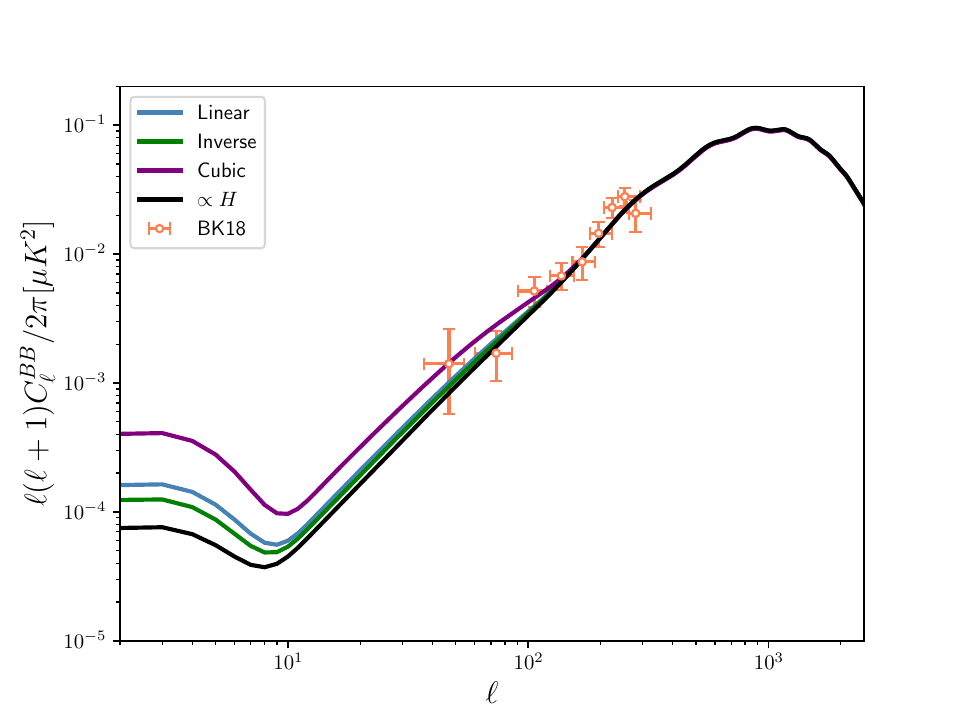}
\caption{CMB B-mode power spectra for the WI models of Table \ref{tab:1}. The data points
correspond to the BK18 data (Fig. 3 of \cite{Santos:2024pix}).}
\label{fig:1}
\end{figure*}

\begin{table*}[t]
\centering
\begin{tabular}{ c c c } 
 \hline
 $\Upsilon = C_C\frac{T^3}{\phi^2}$ & $c=-2,p=3$ & Refs. \cite{Bastero-Gil:2012akf,Berghaus:2019whh} \\ 
 $\Upsilon = C_L T$ & $c=0,p=1$ & Ref. \cite{Bastero-Gil:2016qru} \\ 
 $\Upsilon = C_I\frac{\phi^2}{T}$ & $c=2,p=-1$ & Ref. \cite{Bastero-Gil:2019gao} \\
 $\Upsilon = C_H H$ & $-$ & Refs. \cite{Bolotin:2013jpa,vonMarttens:2019wsc,Barbosa:2017ojt,Motaharfar:2018zyb} \\ 
 \hline
\end{tabular}
\caption{Summary of the models investigated in this work. We always consider the quartic potential given by eq. \ref{eq:2.6}.}
\label{tab:1}
\end{table*}

The findings in \cite{Santos:2024pix} suggest an overall preference for the weak dissipative regime, irrespective of the coefficients considered in Table \ref{tab:1}, with best-fits for the dissipation parameter $\log_{10} Q_\star$ ranging from -1.4 to -2.3. Also, we performed a Bayesian evidence comparison between the quartic WI model with each dissipation coefficient and the standard Starobinsky inflationary model. Following Jeffreys' scale \cite{jeffreys1998theory}, the analyses with all but the linear dissipation coefficient returned a weak to moderate evidence in favor of Starobinsky inflation. However, the test with the linear dissipation case returned an inconclusive evidence, rendering it the most competitive coefficient.

Moreover, the best-fit predictions for the B-mode polarization spectra were provided, which we display here in Fig. \ref{fig:1} for convenience. Although all the models are capable of fitting the B-mode lensing data, there is a clear distinction between the predictions for the primordial B-mode polarization spectrum. This primordial contribution dominates at lower $\ell$, and its amplitude is dependent on the magnitude of tensor perturbations, translated as the tensor-to-scalar ratio $r$. As WI models typically have the characteristic of predicting smaller $r$, it becomes evident that primordial B-modes are particularly sensitive to each dissipation coefficient and future constraints on the tensor-to-scalar ratio are paramount to better distinguish warm inflationary scenarios. This becomes even clearer, as we shall see later, as there are models that predict values of $r$ that might lie outside the future constraints on the parameter. Consequently, if there is no detection of primordial B-modes from the CMB, models that lie in the strong dissipation regime become favored (as $r$ is very low, typically); on the other hand, a positive detection of these modes should exclude many of the potential warm inflationary models in this regime.

\section{Methods and mock likelihoods}\label{sec3}

CMB forecasting has been widely performed over the years \cite{Okamoto:2003zw,Lesgourgues:2005yv,Perotto:2006rj,Smith:2010gu,Wu:2014hta,Errard:2015cxa,Namikawa:2015tba,CORE:2016ymi,Green:2016cjr,SPT:2017ddy,Brinckmann:2018owf,BaleatoLizancos:2020jby,CMB-S4:2020lpa,Hotinli:2021umk,BaleatoLizancos:2022vvr,Hamann:2022apw,Modak:2022gol,Bahr-Kalus:2022prj,Belkner:2023duz,LiteBIRD:2023iei,LiteBIRD:2023aov,LiteBIRD:2023iiy,Chen:2024qlw}, especially with the approval and development of the next generation of CMB dedicated experiments. Surveys scheduled to start operating in this and the next decade aim to achieve precision at cosmological scales not probed before, with the hopes of providing more answers about the cosmos, particularly on the inflationary era. For instance, the \textit{LiteBIRD} satellite \cite{Matsumura:2013aja,Hazumi:2019lys} is expected to reach the large scales of the polarization spectra, in particular those of the primordial B-modes, one of the last pieces of information needed to further support the existence of inflation in the early universe. Complementary to these measurements, ground-based surveys are also crucial at smaller scales. The Simons Observatory \cite{SimonsObservatory:2018koc}, and the CMB-S4 \cite{CMB-S4:2016ple} collaboration, will improve significantly over the \textit{Planck} results, while also giving us clues to the early-time physics.

\subsection{Future CMB surveys}

In what follows, we describe the CMB experiments to be considered in our analysis, focusing on parameters necessary to approach the noise modeling. We briefly describe the nature and status of each experiment:

\begin{itemize}
    \item \textit{LiteBIRD}: The \textit{LiteBIRD} satellite project is being lead by the Japan Aerospace Exploration Agency (JAXA) in collaboration with European and North American countries, with operations scheduled to start by the beginning of the 2030s, and focused on measurements of the CMB polarization modes. It will operate in frequencies between 40 GHz and 402 GHz, divided into 15 bands. We work with the 140 GHz band only, as done in \cite{Brinckmann:2018owf}, and assume a observed sky fraction of $f_{\textrm{sky}}=0.7$, compatible with the current prospects for the experiment \cite{LiteBIRD:2022cnt}.
    
    \item CMB-S4: The collaboration consists of large and small aperture telescopes (LAT and SAT, respectively), projected to be located at both South Pole in Antarctica and the Atacama Desert, in Chile. Being a ground-based experiment, its sky fraction coverage is limited, therefore we assume $f_{\textrm{sky}}=0.4$ ($f_{\textrm{sky}}=0.03$) for LAT (SATs). The experiment is set to cover the frequency range of 30$-$270 GHz.
    
    \item Simons Observatory: Located in Cerro Toco, also in Chile, the Simons Observatory consists of three SATs and one LAT covering frequencies from 27 to 280 GHz, in six bands. We assume in our analysis, the 93 GHz band \footnote{By doing this, we follow the justification presented in \cite{Bahr-Kalus:2022prj}, in which a reasonable approach when considering multiple CMB maps is that the corresponding noise is the minimum of all frequency bands.}, with configurations described in \cite{SimonsObservatory:2018koc}, and a sky fraction of $f_{\textrm{sky}}=0.4$ ($f_{\textrm{sky}}=0.1$), for the LAT (SATs).
\end{itemize}

It is possible to estimate the constraining power of each experiment by assuming different configurations referent to each instrument, such as the sky coverage, temperature and polarization noise, and the scale limit, given by the multipole $\ell$. We use the framework implemented in the \texttt{MontePython} code \cite{Brinckmann:2018cvx,Audren:2012wb}, which builds the mock likelihood given the experimental configuration. Starting from a fiducial model, which is computed with the Cosmic Linear Anisotropy Solving System (\texttt{CLASS}) \cite{Blas:2011rf}, one can either perform a Fisher matrix forecast, or a MCMC analysis around this model, the latter being the case done here. Given the results obtained in \cite{Santos:2024pix}, we choose as a starting model a WI implementation with $\log_{10}Q_\star=-2$, while setting the remaining cosmological parameters with values consistent with \textit{Planck} \cite{Planck:2018jri}, being $\{\omega_b=0.02237,\omega_{cdm}=0.12,100\theta_s=1.04,\ln(10^{10}A_s)=3.045\}$. The optical depth at reionization $\tau_{reio}$, is taken as a Gaussian prior of $\tau=0.0544\pm 0.001$, its uncertainty compatible with the one assumed in \cite{Brinckmann:2018owf}. Table \ref{tab:2} shows the multipole range for each experiment. We perform our uncertainty analysis and plotting of the confidence contours with \texttt{GetDist} \cite{Lewis:2019xzd}. 

\begin{table*}[t]
\centering
\begin{tabular}{ c c c c } 
 \hline 
 Survey & T & E & B \\
 \hline
 SO (LAT) & $30<\ell<8000$ & $30<\ell<8000$ & $30<\ell<1000$ \\ 
 SO (SAT) & $-$ & $30<\ell<8000$ & $30<\ell<300$ \\ 
 CMB-S4 (LAT) & $30<\ell<3000$ & $30<\ell<5000$ & $-$ \\ 
 CMB-S4 (SAT) & $-$ & $21<\ell<335$ & $21<\ell<335$ \\ 
 \textit{LiteBIRD} & $-$ & $2<\ell<200$ & $2<\ell<200$ \\ 
 \hline
\end{tabular}
\caption{Multipole $\ell$ range for the experiments considered in this work.}
\label{tab:2}
\end{table*}

\subsection{Noise modeling and delensing}

In the \texttt{MontePython} implementation, the mock CMB likelihood depends on the $\textbf C_\ell$ matrix, represented as
\begin{equation}
   \textbf C_\ell = \begin{pmatrix}
C_\ell^{TT} + N_\ell^{TT} & C_\ell^{TE} & 0\\
C_\ell^{TE} & C_\ell^{EE} + N_\ell^{EE} & 0\\
0 & 0 & C_\ell^{BB} + N_\ell^{BB}\\ 
\end{pmatrix}\;.
\end{equation}
The noise is included as contributions to the diagonal, modeled as \cite{Ng:1997ez,Wu:2014hta}
\begin{equation}
    N_{\ell}^{XX} = \sigma_X^2\exp\left({\frac{\ell (\ell+1)\theta_{\textrm{\tiny{FWHM}}}^2}{8\ln 2}}\right)\;,
    \label{eq:3.2}
\end{equation}
where $X$ stands for either temperature $T$ or polarization modes $E,B$, representing the sensitivity in units of $\mu$K-arcmin, while $\theta_{\textrm{\tiny{FWHM}}}$ is the beam width, in units of radians. We assume that polarization and temperature noises are related by the multiplicative factor of $\sqrt{2}$, such that $\sigma_{E,B}=\sqrt{2}\sigma_T$, resulting from perfectly polarized detectors in the given experiment.

We follow the same approach as \cite{SimonsObservatory:2018koc}, in which the instrumental noise has a contribution from low frequency contamination at large scales \cite{Maino:2002bv,Donzelli:2009ya}. This $1/f$ additional noise contributes to the instrumental one as
\begin{equation}
    N_{1/f} = 1+\left(\frac{\ell}{\ell_{\textrm{knee}}}\right)^{\alpha_{\textrm{knee}}}.
    \label{eq:3.3}
\end{equation}
The parameters $\ell_{\textrm{knee}}$ and $\alpha_{\textrm{knee}}$ for all experiments considered are the ones displayed in Table 3 of \cite{Bahr-Kalus:2022prj}. For \textit{LiteBIRD}, the beam width and sensitivities are chosen as being correspondent to the 140 GHz band, with values shown in \cite{Hazumi:2019lys,Paoletti:2019pdi}. As for SO, we consider the 93 GHz band of the baseline setting, with parameters displayed in \cite{SimonsObservatory:2018koc}. Finally, for CMB-S4, we follow \cite{Brinckmann:2018owf}, and choose $\theta_{\textrm{\tiny{FWHM}}}=3$ arcmin and $\sigma_T=1 \mu$K-arcmin. The chosen approach involving eqs. \ref{eq:3.2} and \ref{eq:3.3} provides a good approximation to the noise curves shown in Fig. 2 of \cite{SimonsObservatory:2018koc}. We also consider the same multipole range of the experiments as done in \cite{Bahr-Kalus:2022prj}, which are summarized in Table \ref{tab:2}. Note that we consider only polarization measurements from \textit{LiteBIRD}, and avoid the overlap of the $\ell$ for the CMB-S4+\textit{LiteBIRD} and SO+\textit{LiteBIRD} combinations. Furthermore, extra care must be taken when we consider B modes in the analysis. As mentioned, the BB power spectrum is dominated by lensed E modes that mask the contribution from primordial perturbations. A technique known as \textit{delensing} consists of reconstructing the spectrum with reduced lensing power, which depends on the specifics of the instrument and allows us to obtain more information on potential primordial sources at smaller scales. Within the forecasting formalism, one manner of modeling the delensing process is to add an extra source of noise in the $C_\ell$ coefficients, such that $C_\ell \rightarrow C_\ell+ N_{\textrm{del}}$, where $N_{\textrm{del}}$ is the noise associated to the delensing process. An approach to this is to simply estimate the delensing level by a parameter $A_{\textrm{lens}}$, such that the $C_\ell$s with the delensing contribution are written as \cite{Bahr-Kalus:2022prj}
\begin{equation}
    C_\ell^{BB} = A_{\textrm{lens}}\left(C_\ell^{\textrm{BB},\textrm{lensed}}-C_\ell^{\textrm{BB},\textrm{unlensed}}\right) + C_\ell^{\textrm{BB},\textrm{unlensed}},
\end{equation}
plus the instrumental noise term. Note that if $A_{\textrm{lens}}=1$, there is no delensing, while $A_{\textrm{lens}}=0$ corresponds to a perfect delensing. For SO, we use $A_{\textrm{lens}}=0.5$, while $A_{\textrm{lens}}=0.27$ is considered for CMB-S4. As for \textit{LiteBIRD}, we choose a conservative approach and take $A_{\textrm{lens}}=1$ for the experiment.

\begin{table*}[t]
	\centering
	\begin{tabular}{>{\scriptsize}c >{\scriptsize}c >{\scriptsize}c >{\scriptsize}c >{\scriptsize}c }
        \hline
		\hline
		& Cubic coefficient & Linear coefficient & Inverse coefficient & $\propto H$ coefficient \\
		\hline
		\hline 
		\multicolumn{5}{c}{SO}\\
		
		$\sigma(\Omega_b h^2)$   & $0.000058$ &  $0.000055$ & $0.000057$ & $0.0000592$ \\
		$\sigma(\Omega_{c} h^2)$ & $0.00043$ &  $0.00044$ & $0.000445$ & $0.000434$ \\
		$\sigma(100\theta)$ & $0.00012$ &  $0.00012$ &  $0.0000115$ & $0.00012$\\
		$\sigma(\tau)$ & $0.001$ &  $0.001$ &  $0.001$ & $0.001$\\
		$\sigma(\ln(10^{10}A_s))$ & $0.0022$ &  $0.0021$ &  $0.0021$ & $0.0022$\\
		$\sigma(\log_{10}Q_\star)$ & $0.17$ &  $0.253$ &  $0.236$ & $0.174$\\ 
        FoM$/10^{5}$ & $15.759$ & $14.573$ & $14.693$ & $15.411$ \\
		\hline
        \multicolumn{5}{c}{CMB-S4}\\
		
		$\sigma(\Omega_b h^2)$   & $0.000031$ &  $0.000030$  & $0.0000295$ & $0.0000295$\\
		$\sigma(\Omega_{c} h^2)$ & $0.000365$ &  $0.00037$ &  $0.00036$ & $0.00362$\\
		$\sigma(100\theta)$ & $0.000081$ &  $0.000079$ & $0.000079$ & $0.00008$ \\
		$\sigma(\tau)$ & $0.001$ &  $0.001$ &  $0.001$ & $0.001$ \\
		$\sigma(\ln(10^{10}A_s))$ & $0.002$ &  $0.002$ & $0.002$ & $0.002$ \\
		$\sigma(\log_{10}Q_\star)$ & $0.167$ &  $0.223$ &  $0.202$ & $0.206$ \\ 
        FoM$/10^{5}$ & $23.347$ & $23.246$ & $24.312$ & $23.598$ \\ 
        \hline
        \multicolumn{5}{c}{SO+\textit{LiteBIRD}}\\	
		$\sigma(\Omega_b h^2)$   & $0.000056$ &  $0.000054$ &  $0.000054$ & $0.000054$ \\
		$\sigma(\Omega_{c} h^2)$ & $0.00042$ &  $0.00042$ &  $0.00042$ & $0.00043$ \\
		$\sigma(100\theta)$ & $0.00011$ &  $0.00012$ &  $0.00012$ & $0.00012$ \\
		$\sigma(\tau)$ & $0.00092$ &  $0.00091$ &  $0.00094$ & $0.00094$ \\
		$\sigma(\ln(10^{10}A_s))$ & $0.0019$ &  $0.0019$ &  $0.0019$ & $0.002$ \\
		$\sigma(\log_{10}Q_\star)$ & $0.1$ &  $0.16$ &  $0.13$ & $0.13$ \\
        FoM$/10^{5}$ & $20.836$ &  $18.589$ &  $19.207$ & $19.107$ \\
		\hline
        \multicolumn{5}{c}{CMB-S4+\textit{LiteBIRD}}\\	
		$\sigma(\Omega_b h^2)$   & $0.00003$ &  $0.00003$ &  $0.00003$ & $0.00003$ \\
		$\sigma(\Omega_{c} h^2)$ & $0.00036$ &  $0.00037$ &  $0.00037$ & $0.00036$ \\
		$\sigma(100\theta)$ & $0.00008$ &  $0.00008$ &  $0.000083$ & $0.000081$ \\
		$\sigma(\tau)$ & $0.00091$ &  $0.0009$ &  $0.00092$ & $0.00091$ \\
		$\sigma(\ln(10^{10}A_s))$ & $0.0019$ &  $0.0018$ &  $0.0019$ & $0.0018$ \\
		$\sigma(\log_{10}Q_\star)$ & $0.115$ &  $0.165$ &  $0.15$ & $0.155$ \\
        FoM$/10^{5}$ & $29.084$ &  $26.000$ &  $26.419$ & $26.481$ \\
		\hline
		\hline
  
	\end{tabular}
	\caption{ From left to right, the columns show the projected uncertainties at 68\% CL and Figure of Merit values for the cubic, linear, inverse and $\propto H$ dissipation coefficients, according to each experiment.}
	\label{tab:3}
\end{table*}
 
\section{Results}\label{sec4}

In this section, we present our results, with the general consequences for the warm inflationary picture and its comparison with the standard $\Lambda$CDM$+r$ model.

\subsection{Results for WI models}

\begin{figure*}[t]
\centering
\includegraphics[width=\columnwidth]{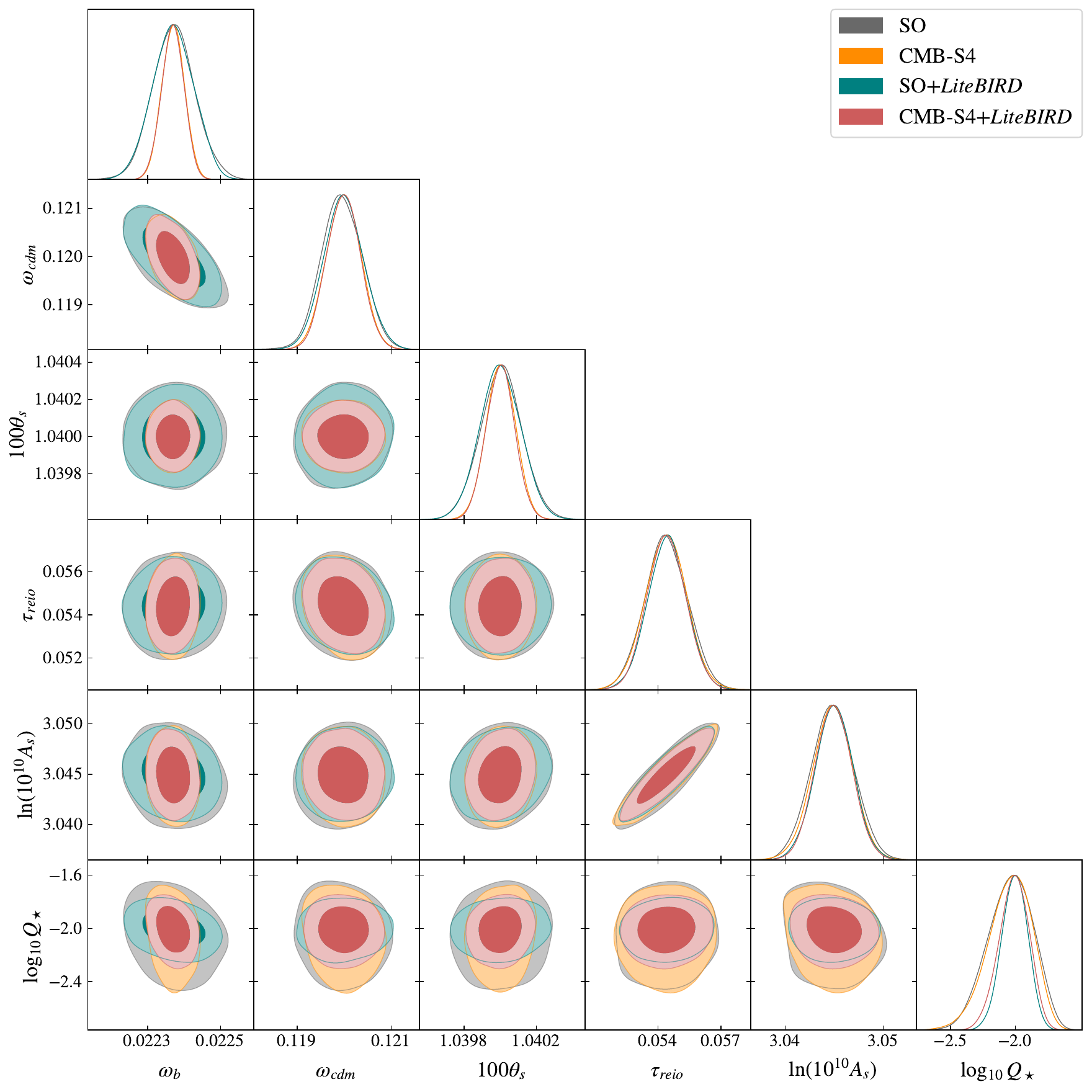}
\caption{Mock confidence contour plots at 68\% and 95\% CL (confidence level) and normalized posteriors for the cubic model. We consider the Simons Observatory (gray contours), CMB-S4 (orange contours), SO+\textit{LiteBIRD} (teal contours), and CMB-S4+\textit{LiteBIRD} (red contours).}
\label{fig:2}
\end{figure*}

\begin{figure*}[t]
\centering
\includegraphics[width=\columnwidth]{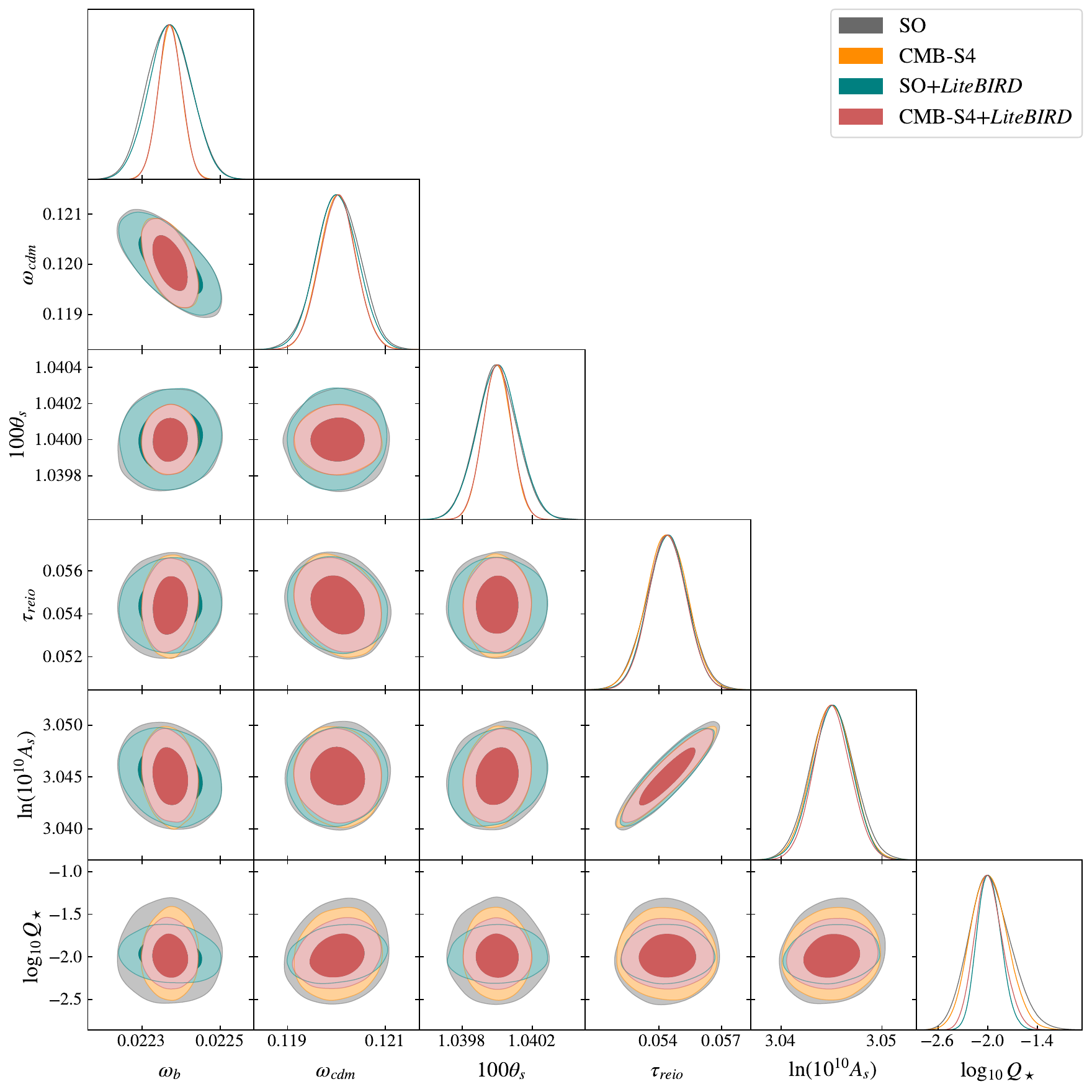}
\caption{Same as fig. \ref{fig:2}, for the linear model.}
\label{fig:3}
\end{figure*}

\begin{figure*}[t]
\centering
\includegraphics[width=\columnwidth]{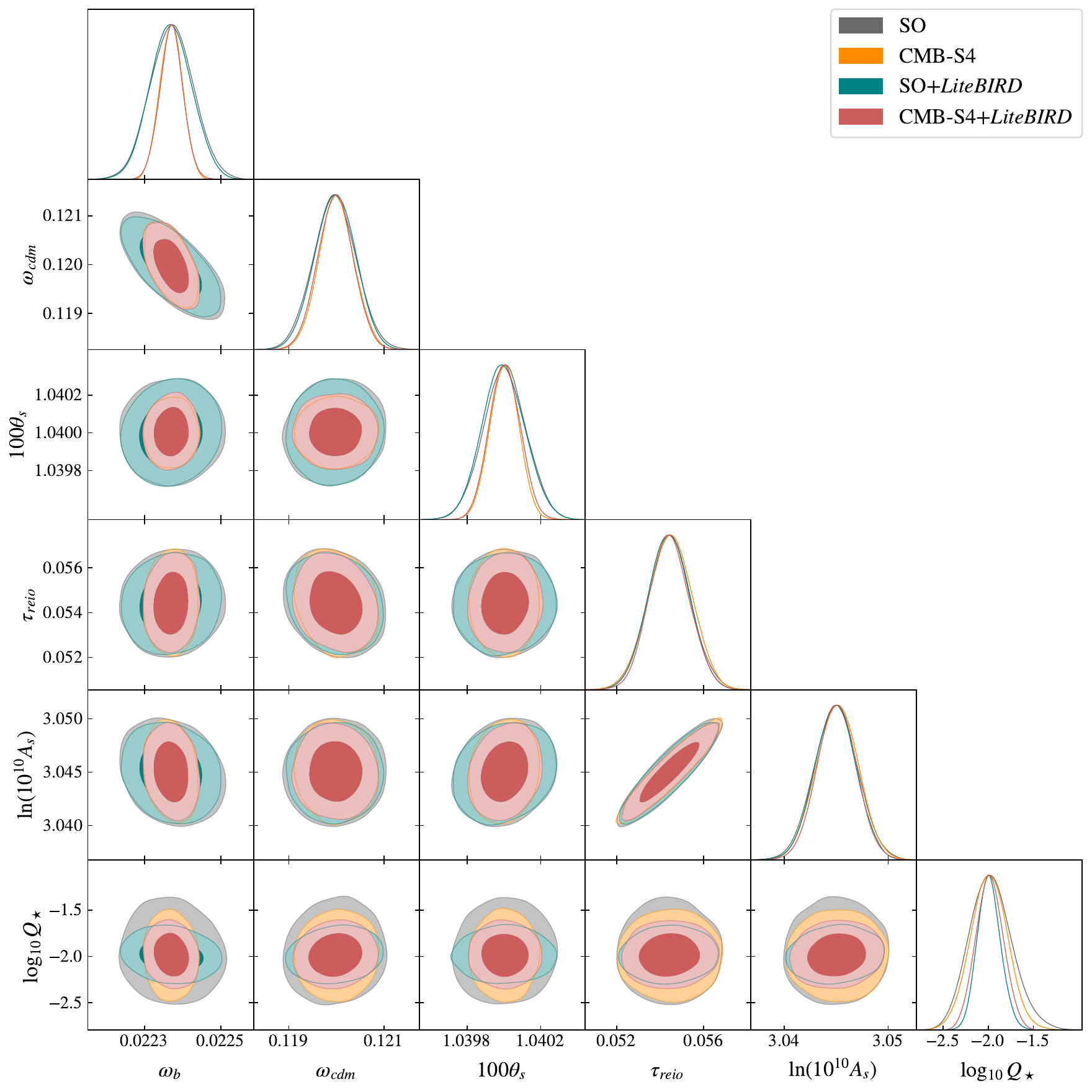}
\caption{Same as fig. \ref{fig:2}, for the inverse model.}
\label{fig:4}
\end{figure*}

\begin{figure*}[t]
\centering
\includegraphics[width=\columnwidth]{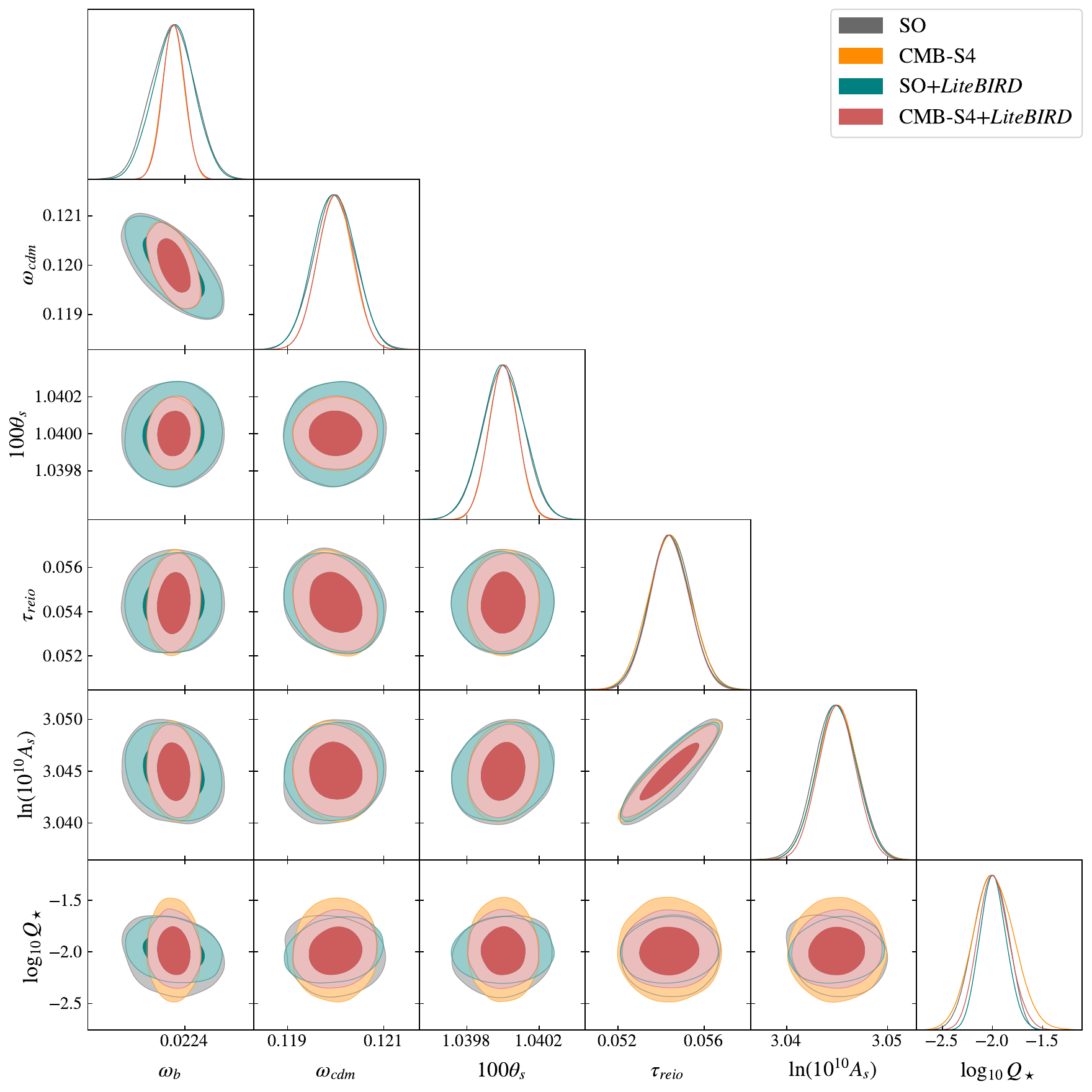}
\caption{Same as fig. \ref{fig:2}, for the $\Upsilon\propto H$ model.}
\label{fig:5}
\end{figure*}

We now discuss our results for the chosen models. Figures \ref{fig:2},\ref{fig:3},\ref{fig:4} and \ref{fig:5} show the contours at 68\% and 95\% CL for the cubic, linear, inverse and $\Upsilon\propto H$ coefficients, respectively. As discussed, we probe the same cosmological parameters as done in \cite{Santos:2024pix} to facilitate the comparison between the estimates. We start by comparing projections from the Simons Observatory and CMB-S4, the two ground-based experiments. Fig. \ref{fig:2} shows the result for the cubic model, where we see a significant improvement in the uncertainties for all parameters. This is explicit if we remember the present estimated value of $\log_{10}Q_\star=-2.32^{+0.50}_{-0.33}$ \cite{Santos:2024pix}. On the other hand, we note that CMB-S4 performs better, especially in estimating the matter fraction, while other parameters remain somewhat similar. It is noticeable here, and this is one of our main results, how the dissipation ratio $Q_\star$ is well determined for either SO or CMB-S4, but in particular for the combination of ground and space-borne experiments, represented by the CMB-S4+\textit{LiteBIRD} and SO+\textit{LiteBIRD} analyses. We symmetrize the uncertainties and show their values in Table \ref{tab:3}. We find $\sigma(\log_{10}Q_\star)=0.115$ and $\sigma(\log_{10}Q_\star)=0.1$ for both experiments, respectively, which is about four times smaller than what was given by the \textit{Planck}+BK18 analysis performed in \cite{Santos:2024pix}. Thus we expect a significant constraining power of future data for these models. 

The results for the other models give us the same general conclusions, in which the uncertainties for each experiment are roughly the same, with the notable exception of $\log_{10}Q_\star$. In the previous CMB analysis, we have seen that while the cubic model has the best constraints on $\log_{10}Q_\star$ (but it is the worst when considering statistical criteria), the other models usually have a larger uncertainty, mainly due to the freedom in the behavior of the primordial power spectrum which is reflected in the curves for the spectral index $n_s$ and tensor-to-scalar ratio $r$. Therefore, it is important to know the extent of constraining power of future CMB surveys in order to determine these parameters, since, depending on the model considered, $Q_\star$ could span a few orders of magnitude, while still providing a viable inflationary picture.

Table \ref{tab:3} shows that the constraints on $\log_{10}Q_\star$ for the linear model have improved by $4.0/4.5$ times when considering SO/CMB-S4, while we have $6.2/6.1$ times when the SO+\textit{LiteBIRD}/CMB-S4+\textit{LiteBIRD} combinations are considered. Even better restrictions are obtained for the inverse model, which has similar uncertainties as the linear model within current CMB data. Moreover, the model where $\Upsilon\propto H$ has some interesting features. First of all, this is the model that allows the widest range in $Q_\star$, where a scenario with a very strong dissipative regime is possible. However, the constraints with SO/CMB-S4 already allow some of the best projections for $\log_{10}Q_\star$ when compared with the other models (losing only to the cubic model), while the SO+\textit{LiteBIRD} and CMB-S4+\textit{LiteBIRD} analyses show that the uncertainty in $\log_{10}Q_\star$ is also decreased to the same level of the other scenarios considered.

\begin{figure*}[t]
\centering
\includegraphics[width=0.7\columnwidth]{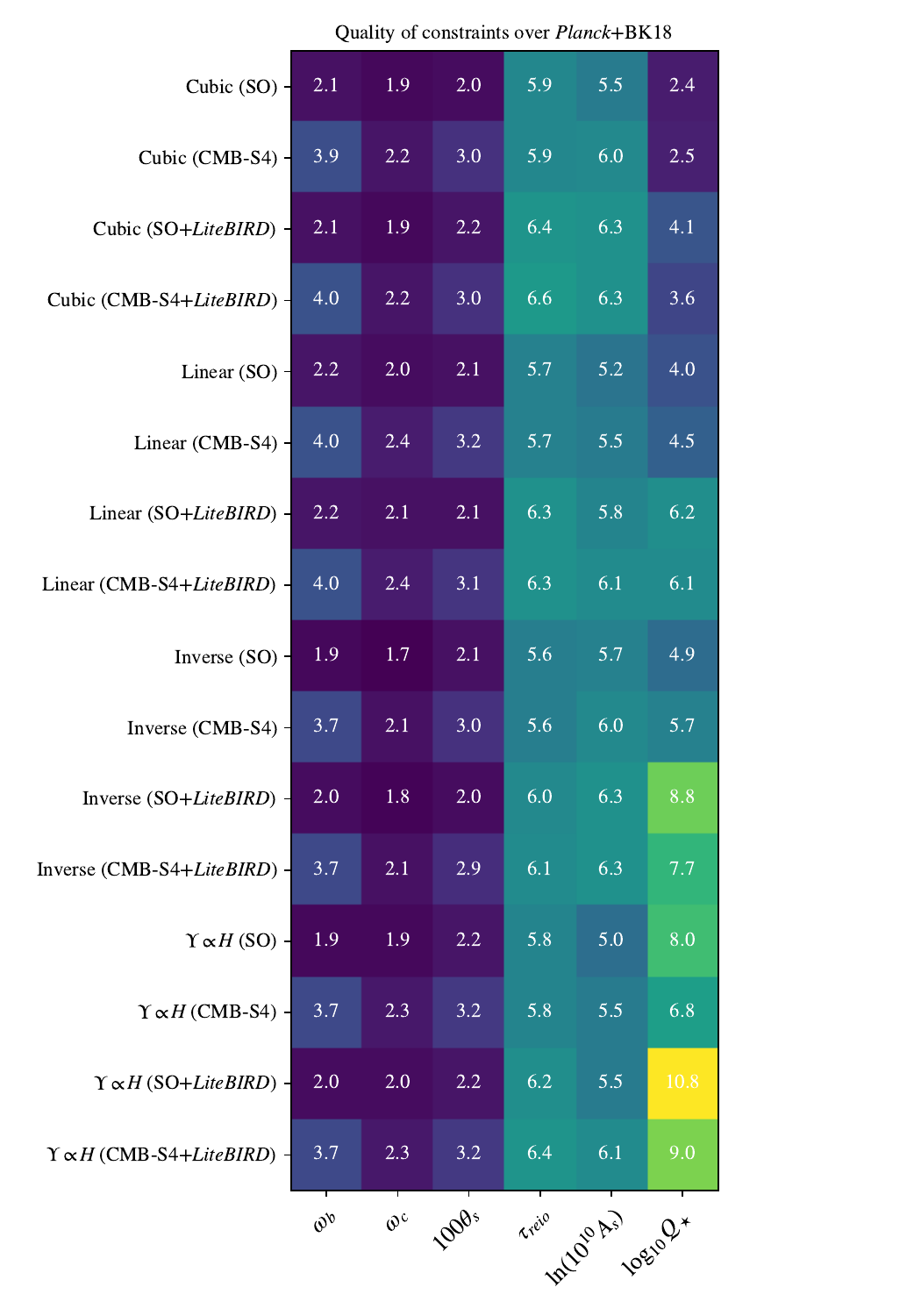}
\caption{The projected quality on the constraining power of future surveys to the WI models, when compared to current CMB data results, presented in \cite{Santos:2024pix}. In the boxes, we show the ratio $\sigma_{\textrm{current}}/\sigma_{\textrm{s4}}$.}
\label{fig:6}
\end{figure*}

We gather the results on the individual improvement in the uncertainties with respect to the current CMB estimates. Figure \ref{fig:6} shows the ratio $\sigma_{\textrm{current}}/\sigma_{\textrm{s4}}$ for each probed model and experiment, for the six parameters chosen. $\sigma_{\textrm{current}}$ is the symmetrized uncertainty obtained from \cite{Santos:2024pix}, while $\sigma_{\textrm{s4}}$ corresponds to the values shown in Table \ref{tab:3}. For all models, Simons Observatory gives an average of two times better constraints on $\omega_b,\omega_c$ and $100\theta_s$, while we note the significant improvement in the dissipation ratio $Q$, represented as $\log_{10}Q_\star$. The cubic model is already the best restricted one, but has an improvement of 2.4 times with respect to \textit{Planck}+BK18 constraints. On the other hand, the $\Upsilon\propto H$ model has a dramatic change, as we estimate a factor of 8 in the uncertainty ratio of the parameter.

A notable change happens when we consider CMB-S4 for all models. $\omega_c, \omega_b$ and $100\theta_s$ constraints are greatly improved when compared to SO, while $\log_{10}Q_\star$, has a milder improvement in comparison, as seen in Table \ref{tab:3}. Now, for CMB-S4+\textit{LiteBIRD}, we note that the uncertainties are improved progressively for the cubic, linear, inverse and $\Upsilon\propto H$ models, respectively. In particular, looking at Figure \ref{fig:6}, constraints for $\log_{10}Q_\star$ in the $\Upsilon\propto H$ model are expected to be $9$ times better than the current CMB estimate, while the SO+\textit{LiteBIRD} analysis gives the best improvement over the current analysis for this parameter, providing a factor of 10.8 for the $\sigma_{\textrm{current}}/\sigma_{\textrm{s4}}$ ratio. Comparing both joint analyses, one can notice that the CMB-S4+\textit{LiteBIRD} combination tends to do better overall than the combination of \textit{LiteBIRD} with SO. This is mainly due to the best restrictions on the other cosmological parameters apart from $\log_{10}Q_\star$; on the other hand, it is noticeable how the combination of both ground-based experiments with \textit{LiteBIRD} expectations should provide equally good constraints in the $\log_{10}Q_\star$ parameter, which is the central one in our implementation of the warm inflationary models.

Another way of indicating the experiment's precision is the computation of the Figure of Merit  (FoM) for each model, which provides the inverse of the area of the ellipse that encloses the 95\% confidence limit. Therefore, a higher FoM value indicates greater accuracy of the experiment for that set of parameters. We can define the FoM for $N$ parameters as \cite{Alonso:2016suf}
\begin{equation}
    \text{FoM} = \left(\det F_{ij}^{-1}\right)^{-1/N},
\end{equation}
where $F_{ij}^{-1}$ is the covariance matrix. As discussed, the most constrained results arise from the CMB-S4+\textit{LiteBIRD} combination. CMB-S4 and the Simons Observatory give us very close constraints, depending on the dissipation coefficient considered. Table~\ref{tab:3} also shows the FoM measurements for the different experiments and models. One can notice that for the Simons Observatory, the better accuracy is obtained when considering the cubic dissipation coefficient, with FoM$/10^{5}=15.759$ while CMB-S4 gives FoM$/10^{5}=24.312$ for the inverse model as the highest value. When we look at the CMB-S4+\textit{LiteBIRD} combination, we notice the cubic coefficient as the one that provides the best constraints. For this case, we have FoM$/10^{5}=29.084$, while the SO+\textit{LiteBIRD} analysis gives FoM$/10^{5}=20.836$.

\subsection{Comparing WI models within a $r=0$ compatibility}

\begin{figure*}[t]
\centering
\includegraphics[width=0.7\columnwidth]{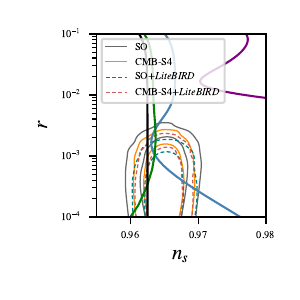}
\caption{Projected constraints on the scalar spectral index $n_s$ and tensor-to-scalar ratio $r$ for the $\Lambda$CDM$+r$ model, for SO (gray contours), CMB-S4 (orange contours), SO+\textit{LiteBIRD} (dashed teal contours) and CMB-S4+\textit{LiteBIRD} (dashed red contours), where we show the 68\% and 95\% CL constraints. Also in the plot are the predictions for the different WI models investigated in this work, being the cubic (purple curve), linear (blue curve), inverse (green curve), and the $\Upsilon\propto H$ model, represented by the black curve.}
\label{fig:7}
\end{figure*}

For completeness, we also present the prospects for warm inflation in light of the standard $\Lambda$CDM$+r$ model. Current observations are compatible with $r=0$, and it is possible that these results could be maintained, with the upper limit being decreased. We perform a forecast for this model, assuming a $r=0$ detection, with results shown in Fig. \ref{fig:7}, for the same mock likelihoods previously discussed. In the figure, we show the predicted $n_s-r$ contours for each experiment, compared with the predictions for each WI model.

Regardless of the choice of survey, the cubic WI model (purple curve) becomes excluded by the improved constraints on $r$; this not only shows how warm inflationary models can be now distinguished even better, but also tells us how in the event of a non-detection of a primordial gravitational wave signal, possibly dozens of other inflationary models will also be excluded by data as well. On the other hand, such scenario is somewhat more favorable to WI models, which allow in general a decreasing tensor-to-scalar ratio with respect to cold inflationary models. The other WI models presented are still in good agreement with a lower $r$, as also seen in the figure. For instance, the linear model, shown by the blue curve in the plot, is not only favored statistically with respect to the Starobinsky model \cite{Santos:2024pix}, but is also the one that better agrees with the $1\sigma$ contour of the SO+\textit{LiteBIRD}/CMB-S4+\textit{LiteBIRD} projections. Also, viable WI models in the very strong regime ($Q\gg1$) start to become stronger candidates as well, since they generally predict extremely low values of $r$.

\section{Discussion and Conclusions}\label{sec5}

The next decade CMB data is highly anticipated, not only because it will open a new era in precision cosmology, after the successful findings of \textit{WMAP} and \textit{Planck}, but also for the prospects of further exploring the parameter space of a variety of early-time physical models, including inflation. As there are plenty of viable models currently, it becomes urgent to start properly distinguishing these scenarios, as well as looking for specific phenomena directly related to a possible inflationary period. The most searched signal from primordial times is arguably the production of gravitational waves during inflation, which manifest themselves through a polarization pattern in the CMB. If existent, these B modes are actually extremely difficult to detect, due to the weakness of the signal, and the astrophysical contaminants that should be properly extracted in the data processing. Thus, it is important to assess the readiness of developing experiments in performing these measurements, while also dealing with different kinds of noise that will permeate the observations.

In this work, we performed a forecast to investigate the sensitivity with which future CMB experiments will constrain early-time physics within the warm inflation scenario, which has been shown to be promising due to a solid theoretical background, rich phenomenology, and the possibility of uniting the eras of inflation and reheating in the same framework. Specifically, we explore a model given by a quartic potential, considering four different forms of the dissipation coefficient. We then have forecast restrictions on these models using configurations of Simons Observatory, CMB-S4, and combinations of CMB-S4 and SO with \textit{LiteBIRD}. It is important to notice that more accurate prospects for each of these experiments are currently in development, and important issues such as delensing and the proper treatment of foregrounds \cite{Beck:2020dhe,LiteBIRD:2023aov,SimonsObservatory:2024gol} should affect in a significant manner possible constraints. In our approach, while we provide an optimistic perspective for the characterization of WI models, an important finding here is the factor of improvement over the current CMB constraints obtained in \cite{Santos:2024pix}, something that might give a good perspective of characterizing these models in the future.

A summary of the main findings of this work is displayed in Table \ref{tab:3}, which shows the projected uncertainties in the cosmological parameters, and Figure \ref{fig:6}, in which the improvements over current CMB constraints are shown. The improvement in the uncertainties is compatible with the findings in \cite{Brinckmann:2018owf}, for instance, where a stage IV CMB forecast was also performed, but for an extended $\Lambda$CDM model with neutrinos. We see how especially the combination of CMB-S4 and \textit{LiteBIRD} could lead to at least $\sim$2 times smaller uncertainties with respect to current data. As seen in Figure \ref{fig:6}, the consequences are even more explicit for the logarithm of the dissipation ratio $\log_{10}Q_\star$, which has its uncertainties reduced by at least 2.4 times. Another important discussion is the possibility of a non-detection of a primordial gravitational wave signal, leading to results compatible with $r=0$, which is the case currently. We perform an analysis for the $\Lambda$CDM$+r$ model, and compare the $n_s-r$ contours with the predictions for each WI model. Since we have around $r<3\times 10^{-3}$ at 95\% CL, the cubic WI model becomes excluded; this tells us that future CMB data should allow us to distinguish different dissipation coefficients already at the $n_s-r$ level. Moreover, the linear model with a quartic potential, compatible with the proposal in \cite{Bastero-Gil:2016qru}, becomes a even more viable option in the WI construction. Finally, as mentioned, a $r=0$ result can be favorable to WI models in the strong dissipation regime, in which the tensor-to-scalar ratio is typically low, and few well motivated cold inflationary models are actually compatible with such projections.

\section*{Acknowledgments}

FBMS is supported by Conselho Nacional de Desenvolvimento Científico e Tecnológico (CNPq) grant No. 151554/2024-2. GR and RdS are supported by the Coordena\c{c}\~ao de Aperfei\c{c}oamento de Pessoal de N\'ivel Superior (CAPES). JSA is supported by CNPq grant No. 307683/2022-2 and Funda\c{c}\~ao de Amparo \`a Pesquisa do Estado do Rio de Janeiro (FAPERJ) grant No. 259610 (2021). We also acknowledge the use of the \texttt{MontePython}, \texttt{CLASS} and \texttt{GetDist}. This work was developed thanks to the use of the National Observatory Data Center (CPDON).

\bibliography{references}

\end{document}